\documentclass[aps,prl,reprint,amsmath,amssymb,showpacs,superscriptaddress]{revtex4-2}
\usepackage{CJK}
\usepackage{graphicx}
\usepackage{dcolumn}
\usepackage{bm}
\usepackage{tabularx}
\usepackage{multirow}
\usepackage{lettrine}
\usepackage{color}
\usepackage{booktabs}

\begin{document}
\title{Dynamical Triaxial Correlations in the Charge Radii of Neutron-Rich Ruthenium Isotopes}

\author{Y. K. Wang}
\email[]{wangyk@buaa.edu.cn}
\affiliation{School of Physics, Beihang University, Beijing 102206, China}

\date{\today}
\begin{abstract}
Nuclear charge radii are fundamental observables that track changes in nuclear structure. 
Recent measurements of neutron-rich Ru isotopes, interpreted using mean-field density-functional calculations, have highlighted the sensitivity of charge radii to static triaxial deformation. 
Whether dynamical shape fluctuations involving triaxial configurations---known to affect both nuclear ground- and excited-state properties---can reshape the charge-radii evolution remains unresolved.
In this Letter, we investigate the charge radii of even-even neutron-rich $^{106-114}$Ru within a nuclear density-functional framework that combines three-dimensional angular-momentum projection with the generator-coordinate method.
Including dynamical triaxial shape fluctuations yields charge radii in good agreement with experiment, without introducing additional adjustable parameters.
By successively incorporating static triaxial deformation, dynamical axial shape fluctuations, and dynamical triaxial shape fluctuations, we trace how the predicted charge radii change at each step. 
These results demonstrate that triaxial shape dynamics can leave a substantial imprint on charge radii. 
They also inform the interpretation of experimental charge-radius measurements used to trace the evolution of nuclear structure, since charge radii alone cannot uniquely determine the underlying nuclear shape.

\end{abstract}
\maketitle
\date{today}

%------------------------------------------------------------------------
\textit{Introduction}--- Nuclear charge radii provide a precise probe of evolving nuclear structure~\cite{Gorges:2019wzy,Marsh:2018wxs,Barzakh:2021gfl,Yang:2016tyv,Hagen:2015yea,Kreim:2013uqb}, yet their relationship to triaxial deformation is subtle. 
Because the root-mean-square charge radius depends on the angle-averaged charge density, it cannot, by itself, uniquely determine the intrinsic triaxial shape. 
In a volume-conserving liquid-drop picture, the leading quadrupole correction to the mean-square radius depends on the magnitude of deformation through $\beta$, but not on the triaxiality parameter $\gamma$~\cite{BohrMottelson1975}. 
This leading-order independence from $\gamma$ need not hold in a microscopic description: varying $\gamma$ can rearrange the single-particle spectrum~\cite{Niksic:2011sg} and thereby change the charge density.
Indeed, recent collinear laser spectroscopy measurements of neutron-rich $^{106-114}$Ru have yielded charge radii that are described better by density-functional calculations allowing triaxial mean-field shapes than by calculations restricted to axial symmetry~\cite{Maass:2025gnq}.
These findings demonstrate the relevance of static triaxiality within the microscopic framework but leave open how dynamical triaxial correlations affect the radii of the correlated nuclear states. 

The dynamical triaxial correlations are especially important for nuclei with soft potential-energy surfaces (PESs), where the ground-state wave function can extend over a range of intrinsic shapes. 
Shape fluctuations have been demonstrated to play crucial roles in a wide range of nuclear phenomena, spanning from nuclear mass~\cite{Yang:2021wcd}, shape coexistence~\cite{Yang:2023dql,Egido:2020bql,Rodriguez:2011hb}, shape transitions~\cite{Nomura:2017ilh,Li:2010qu,Togashi:2016yzs}, low-lying spectroscopic properties~\cite{Chen:2017vtf,Rodriguez:2010by,Xiang:2023eer}, to nuclear weak decays~\cite{Jiao:2017opc,Wang:2023hkc}.
They have been also shown to affect charge radii, including in calculations that treat the full $(\beta,\gamma)$ collective space within the Gaussian overlap approximation~\cite{Libert:2007mrz}. 
The predicted triaxial softness of neutron-rich Ru \cite{NuclearMap,Esmaylzadeh:2022xgi,Garrett:2020ded,Nomura:2016nyh} and the newly available charge-radius data~\cite{Maass:2025gnq} provide an opportunity to examine a broader issue: whether quantum fluctuations among triaxial shapes can change a scalar ground-state observable beyond the effect of a static triaxial minimum.

In this Letter, we investigate the even-even $^{106-114}$Ru isotopes using a relativistic energy density functional theory combined with three-dimensional angular-momentum projection and generator-coordinate method (RDFT-PGCM). 
Comparing axial and triaxial calculations at the mean-field and projected configuration-mixing levels allows us to disentangle static deformation from dynamical shape correlations. 
It is found that axial shape mixing generally lowers the calculated charge radii, whereas triaxial mixing increases them and brings the results into close agreement with experiment. 
We further analyze the collective shape distributions and the radii of individually projected configurations to identify how these correlations produce the observed change.

\textit{Methods}---The correlated nuclear states are described within the RDFT-PGCM.
The corresponding many-body wave functions are expressed as
\begin{align}
    |\Psi^{IM}_\sigma\rangle
    = \sum_{Kq} F^{I\sigma}_{Kq}
    \hat{P}^I_{MK}|\Phi(q)\rangle,
\end{align}
where $q \equiv (\beta,\gamma)$ denotes the quadrupole deformation parameters, $\hat{P}^I_{MK}$ is the three-dimensional angular-momentum projection operator~\cite{Ring:2004}, and $\sigma$ labels different many-body states with the same angular momentum $I$.
The intrinsic mean-field states $|\Phi(q)\rangle$ are obtained by solving the constrained triaxial relativistic Hartree-Bogoliubov (TRHB) equations~\cite{Meng2016,Niksic:2011sg}, which incorporate both quadrupole deformations and pairing correlations. 
By varying the constraints on $(\beta,\gamma)$, a set of intrinsic configurations spanning the collective deformation space is generated.
Angular-momentum projection then restores the rotational
symmetry broken at the mean-field level, yielding the projected basis states $\{\hat{P}^I_{MK}|\Phi(q)\rangle\}$ with well-defined angular momentum.

The expansion coefficients $F^{I\sigma}_{Kq}$ are determined by diagonalizing the many-body Hamiltonian $\hat{H}$ in the space spanned by these projected basis states.
This leads to the Hill-Wheeler equation~\cite{Yao:2009ry,Egido:2016bdz}
\begin{align}\label{eq:Hill-Wheeler}
    \sum_{K'q'}
    \left[
    \mathcal{H}^I_{Kq;K'q'}
    - E^{I\sigma}
    \mathcal{N}^I_{Kq;K'q'}
    \right]
    F^{I\sigma}_{K'q'} = 0,
\end{align}
where
\begin{align}
    \mathcal{H}^I_{Kq;K'q'}
    & = \langle\Phi(q)|
    \hat{H}\hat{P}^I_{KK'}
    |\Phi(q')\rangle,
    \notag \\
    \mathcal{N}^I_{Kq;K'q'}
    & = \langle\Phi(q)|
    \hat{P}^I_{KK'}
    |\Phi(q')\rangle
\end{align}
are the Hamiltonian and norm kernels, respectively.
These kernels are evaluated using the algorithms developed in Refs.~\cite{Hu:2013dha,Carlsson:2020fiv}.
The resulting generalized eigenvalue problem is solved using standard numerical techniques~\cite{Egido:2016bdz}, yielding the energies $E^{I\sigma}$ and the expansion coefficients $F^{I\sigma}_{Kq}$, which fully specify the correlated many-body wave functions $|\Psi^{IM}_\sigma\rangle$.

With the obtained many-body wave functions, the nuclear root-mean-square charge radii $R_c^{I\sigma}$ are calculated as
\begin{align}
    R_c^{I\sigma}
    = \sqrt{\langle \widehat{r^2}\rangle_{I\sigma}
    + 0.64~\mathrm{fm}^2},
\end{align}
where the constant $0.64~\mathrm{fm}^2$ accounts for the finite size of the proton~\cite{Niksic:2014dra,Bender:2005ri}.
The mean-square radius of the point-proton distribution is defined by
\begin{align}
    \langle \widehat{r^2}\rangle_{I\sigma}
    = \frac{1}{Z}
    \langle\Psi^{I}_\sigma|
    \sum_{i=1}^{A}
    \frac{1-\tau_{3i}}{2}r_i^2
    |\Psi^{I}_\sigma\rangle,
\end{align}
where $Z$ and $A$ denote the proton and mass numbers, respectively, and $\tau_{3i}$ is the third component of the isospin operator for nucleon $i$, with the convention $\tau_{3i} = -1$ for protons and $\tau_{3i} = +1$ for neutrons.

To help interpret the effect of shape mixing on the calculated charge radii, it is necessary to examine how the correlated state is distributed over the $(\beta,\gamma)$ plane.
The expansion coefficients $F^{I\sigma}_{Kq}$, however, do not directly provide such a distribution because the projected basis states are nonorthogonal.
We therefore introduce the collective wave functions~\cite{Ring:2004}
\begin{align}
    G^{I\sigma}_{Kq}
    = \sum_{K'q'}
    (\mathcal{N}^I)^{1/2}_{Kq;K'q'}
    F^{I\sigma}_{K'q'},
\end{align}
with which one can define the following quantity
\begin{align}
    g^{I\sigma}(q)
    = \sum_K |G^{I\sigma}_{Kq}|^2.
\end{align}
It is straightforward to verify that for normalized many-body states satisfying $\langle\Psi^{IM}_\sigma|\Psi^{IM}_\sigma\rangle = 1$, the distribution of $g^{I=0}(q)$ over the $(\beta,\gamma)$ deformation plane satisfies
\begin{align}
    \sum_q g^{I\sigma}(q) = 1.
\end{align}
Accordingly, $g^{I\sigma}(q)$ provides a quantitative measure of the relative weight of different quadrupole deformations in the correlated nuclear states.

In the numerical calculations, the particle-hole part of the many-body Hamiltonian $\hat{H}$ is derived from the nonlinear relativistic point-coupling Lagrangian underlying PC-PK1~\cite{Zhao:2010hi}, while the particle-particle part is described by a finite-range separable pairing interaction~\cite{Tian:2009zzh}. 
The constrained TRHB equations are obtained self-consistently from $\hat H$ at the mean-field level and solved in a three-dimensional harmonic-oscillator basis comprising ten major shells. 
For each quadrupole deformation $q= (\beta,\gamma)$, the constrained solution provides an intrinsic state $|\Phi(q)\rangle$.
For the five even-even isotopes $^{106,108,110,112,114}$Ru, we consider $0\leq \beta \leq0.8$ and $0^\circ \leq \gamma \leq 60^\circ$. 
The deformation space is sampled using the triangular-mesh construction of Ref.~\cite{Rodriguez:2010by}, with a deformation step of $\Delta \beta = 0.1$, yielding 45 intrinsic configurations for each isotope. 
The Euler-angle integrals required for angular-momentum projection are evaluated on a 2016-point grid in $\Omega=(\psi,\theta,\phi)$.
Particle-number projection is not performed. 
To control the average particle numbers in the configuration-mixing calculation, we include the auxiliary term $-\lambda_p (\hat{Z} - Z)-\lambda_n (\hat{N} - N)$ in the Hamiltonian kernels entering the Hill–Wheeler equation~\eqref{eq:Hill-Wheeler}, as suggested in Refs.~\cite{Bonche:1990ghh,Hara:1982egw}.
For each isotope, $\lambda_p$ and $\lambda_n$ are fixed to the averages of the corresponding proton and neutron Fermi energies over the constrained TRHB configurations included in the calculation. 
The resulting ground-state expectation values of the proton number for $^{106}$Ru, $^{108}$Ru, $^{110}$Ru,$^{112}$Ru, and $^{114}$Ru are 43.988, 43.982, 43.986, 43.991, and 44.007, respectively. 
The largest relative deviation from $Z = 44$ is 0.04\%. 
We have verified that the charge radii reported below are converged with respect to the adopted numerical settings. 
No parameters were adjusted specifically to reproduce the Ru charge radii.

%------------------------------------------------------------------
\begin{figure*}[!htbp]
  \centering
  \includegraphics[width=\textwidth]{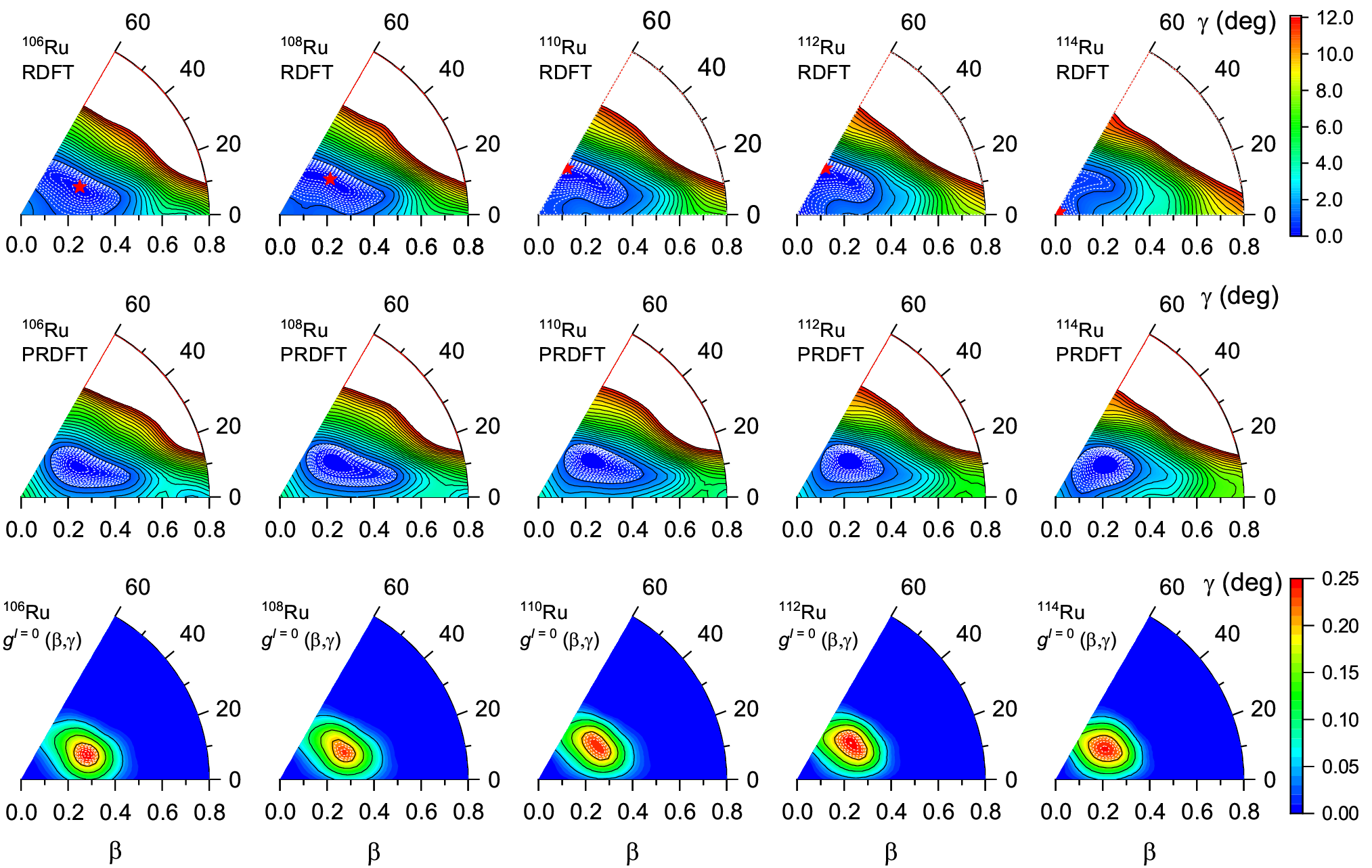}
  \caption{(Color online) PESs and collective shape distributions for $^{106,108,110,112,114}$Ru. 
  The top row shows the mean-field PESs obtained from constrained TRHB calculations; red stars mark their minima. 
  The middle row shows the $I=0$ projected energy surfaces obtained by angular-momentum projection of each intrinsic state, without shape mixing. 
  In both rows, energies are measured relative to the minimum of the corresponding surface. 
  Solid black (dashed white) contours are separated by 0.5 (0.1) MeV. 
  The bottom row shows the collective probability distributions $g^{I=0}(\beta,\gamma)$ of the PGCM ground states, obtained by mixing projected axial and triaxial configurations. 
  Solid black (dashed white) contours in these panels are separated by 0.05 (0.01).
  }
  \label{Fig:PES}
\end{figure*}
%------------------------------------------------------------------------------

\textit{Results}---Figure~\ref{Fig:PES} compares the mean-field PESs obtained by the constrained TRHB calculations, the $I = 0$ surfaces obtained by angular-momentum projection of individual mean-field states without shape mixing, and the collective probability distributions $g^{I=0}(\beta,\gamma)$ obtained from the full PGCM calculations. 
The mean-field PESs are $\gamma$-soft throughout the five even-even Ru isotopes. 
Shallow triaxial minima occur at $\gamma \approx 25^\circ$ and $35^\circ$ in $^{106}$Ru and $^{108}$Ru, respectively. 
In $^{110}$Ru and $^{112}$Ru, the minima shift to $\gamma = 60^\circ$, corresponding to oblate configurations with smaller $\beta$ values than in the lighter isotopes. 
The mean-field minimum becomes spherical in $^{114}$Ru, although an oblate local minimum near $(\beta,\gamma)=(0.2,60^\circ)$ lies only about 0.7 MeV higher. 
These soft surfaces and closely competing configurations motivate an examination of correlations beyond the static mean field.

Indeed, angular-momentum projection substantially changes the mean-field picture. 
As shown in the middle row, the minima of the $I = 0$ projected surfaces lie in the triaxial region for all five isotopes, even before shape mixing is introduced. 
Thus, rotational-symmetry restoration changes the relative energies of competing shapes. 
The projected surfaces remain soft, motivating the inclusion of shape mixing across both the $\beta$ and $\gamma$ directions.

The bottom row shows the collective distributions $g^{I=0}(\beta,\gamma)$ of the PGCM ground states. 
Their weights extend over a range of deformations rather than being concentrated at a single configuration, and their maxima lie at triaxial shapes in all five isotopes. 
The contrast with the mean-field minima is particularly pronounced in $^{110,112,114}$Ru.
Although their mean-field minima are oblate or spherical, triaxial configurations carry the largest collective weights in the correlated ground states. 
The consequences for their charge radii are examined below.

%------------------------------------------------------------------
\begin{figure}[!htbp]
  \centering
  \includegraphics[width=0.48\textwidth]{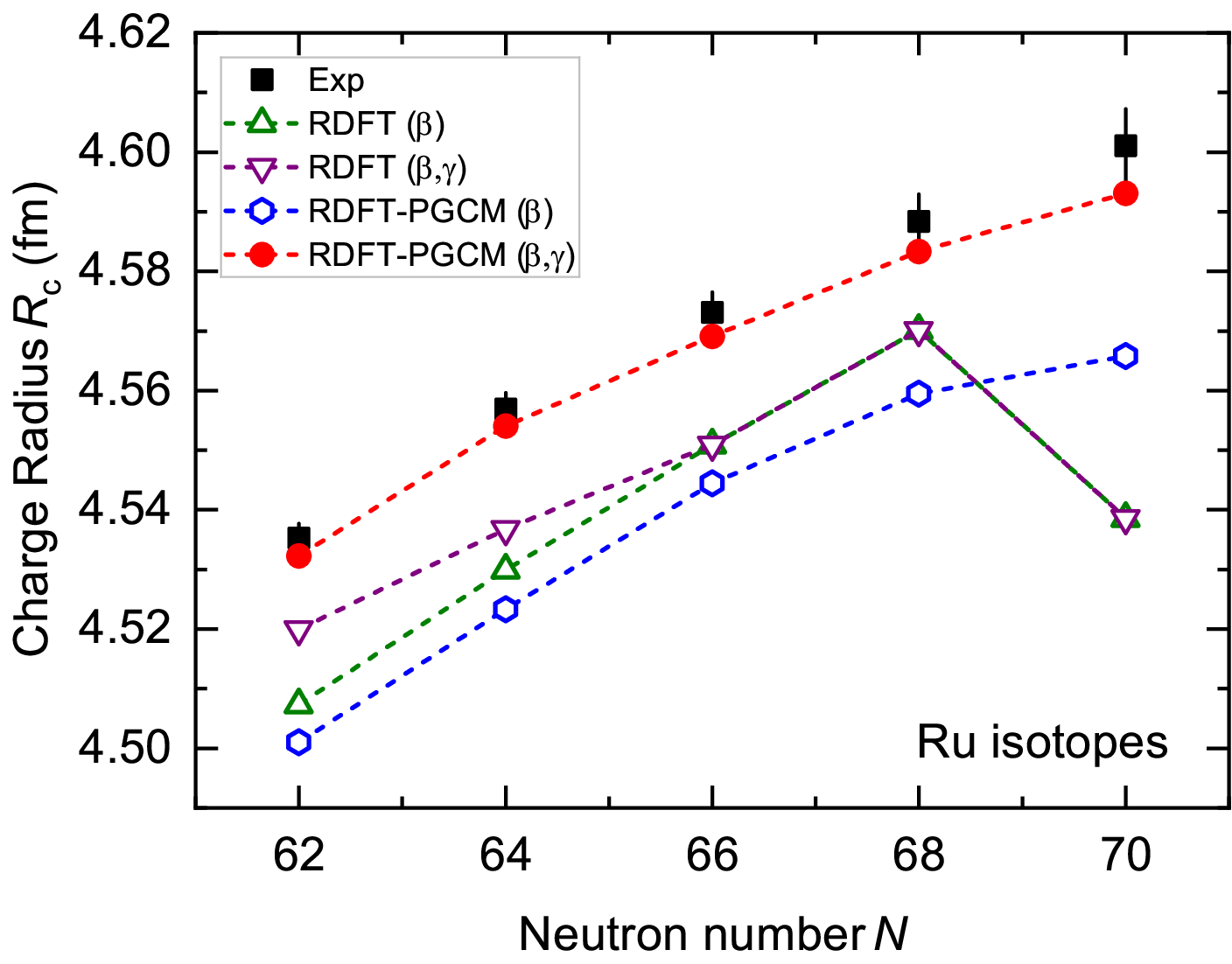}
  \caption{(Color online) Ground-state charge radii of even-even $^{106-114}$Ru isotopes as a function of neutron number $N$.
  Mean-field results restricted to axial symmetry [RDFT ($\beta$)] or allowing triaxial deformation [RDFT ($\beta,\gamma)$] are compared with angular-momentum-projected generator-coordinate results that mix only axially symmetric configurations [RDFT-PGCM ($\beta$)] or configurations across the ($\beta,\gamma$) plane [RDFT-PGCM ($\beta,\gamma$)].
  Experimental values are taken from Ref.~\cite{Maass:2025gnq}. 
  The dashed lines are used to guide the eye.
  }
  \label{Fig:Radii}
\end{figure}
%------------------------------------------------------------------------------

Figure~\ref{Fig:Radii} compares the calculated ground-state charge radii with the experimental values. 
The mean-field calculations restricted to axially symmetric shapes systematically underestimate the data. 
They also predict a sharp decrease from $N=68$ to $N=70$, where the mean-field minimum of $^{114}$Ru becomes spherical, as shown in the top row of Fig.~\ref{Fig:PES}. 
Allowing static triaxial shapes increases the radii of $^{106,108}$Ru, whose mean-field minima are triaxial, and brings them closer to experiment. 
For $^{110,112,114}$Ru, the lowest-energy mean-field configurations remain oblate or spherical, and the radii are essentially unchanged. 
Static triaxiality thus improves the description of the lighter isotopes but does not account for the measured isotopic trend.

Angular-momentum projection and shape mixing restricted to axially symmetric configurations reduce the radii of $^{106,108,110,112}$Ru relative to the corresponding axial mean-field results, increasing their discrepancy with experiment. 
The exception is $^{114}$Ru, for which the axial PGCM radius increases but remains below the measured value. 
Extending the mixing to projected configurations across the $(\beta,\gamma)$ plane raises the radii of all five isotopes relative to the axial PGCM results and brings both their magnitudes and neutron-number dependence much closer to experiment. 
Since both PGCM calculations include angular-momentum projection, their comparison highlights the importance of including triaxial configurations in the correlated ground states.

%------------------------------------------------------------------
\begin{figure}[!htbp]
  \centering
  \includegraphics[width=0.48\textwidth]{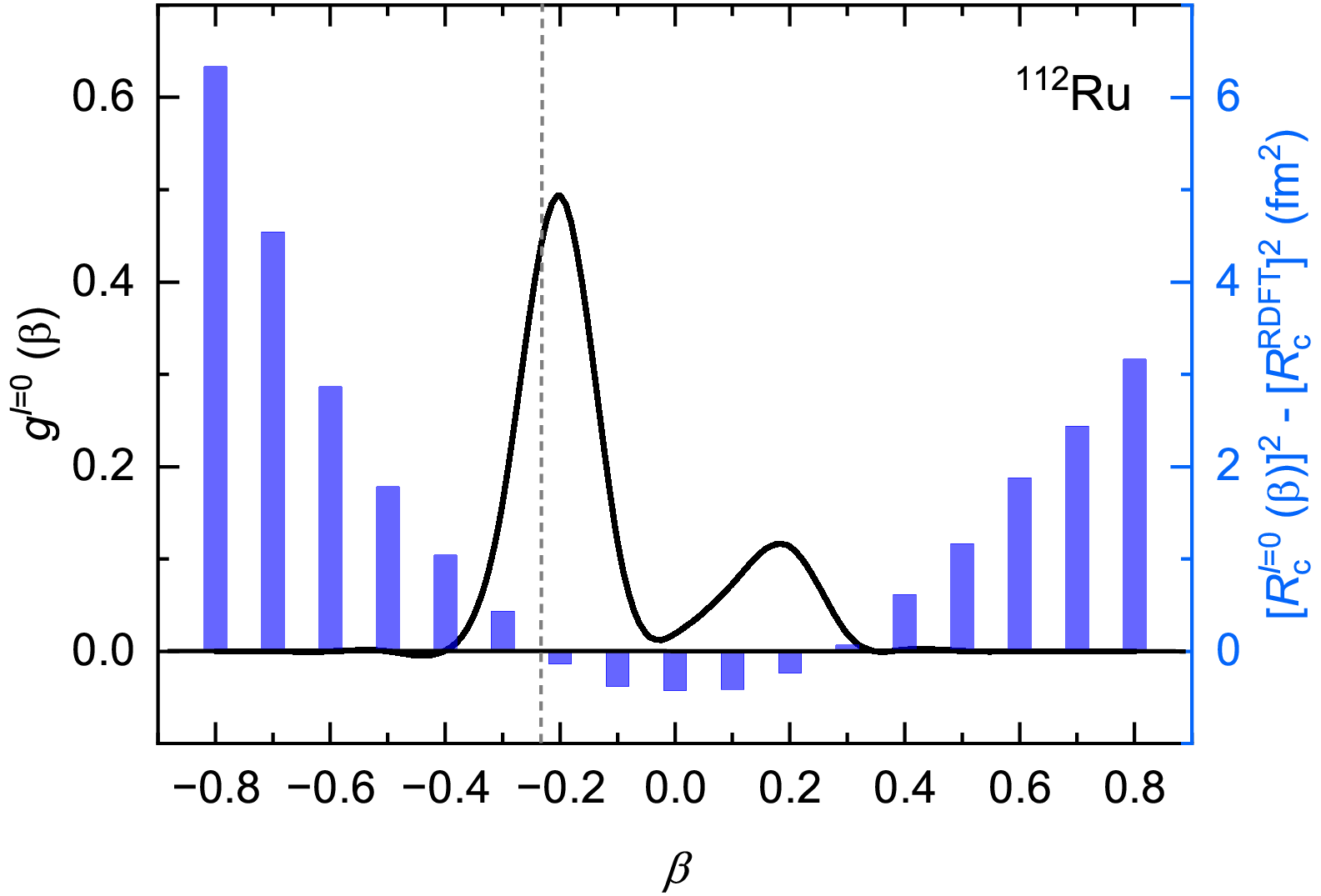}
  \caption{(Color online) Collective probability distribution $g^{I=0}(\beta)$ for the $^{112}$Ru ground state obtained from the axial PGCM calculation (black curve, left axis), together with the difference in squared charge radii, $[R_c^{I=0}(\beta)]^2-[R_c^{\mathrm{RDFT}}]^2$ (blue bars, right axis). 
  Here, $R_c^{I=0}(\beta)$ is the charge radius of an individual axially symmetric configuration after $I=0$ projection without shape mixing, whereas $R_c^{\mathrm{RDFT}}$ is the radius of the lowest-energy mean-field configuration. 
  The vertical gray dashed line marks the deformation of that mean-field minimum. 
  Negative (positive) $\beta$ denotes oblate (prolate) deformation.
  }
  \label{Fig:Axial}
\end{figure}
%------------------------------------------------------------------------------

To understand why the axial PGCM calculations yield smaller charge radii than the mean-field calculations for $^{106,108,110,112}$Ru, we examine $^{112}$Ru as a representative case in Fig.~\ref{Fig:Axial}. 
The figure shows the collective probability distribution $g^{I=0}(\beta)$ together with the difference $\Delta R_{\mathrm{c}}^2(\beta) \equiv [R_{\mathrm{c}}^{I=0}(\beta)]^2-[R_{\mathrm{c}}^{\mathrm{RDFT}}]^2$, where $R_\mathrm{c}^{I=0}(\beta)$ is the radius of an individual axially symmetric configuration after $I=0$ projection, without shape mixing, and $R_\mathrm{c}^{\mathrm{RDFT}}$ refers to the lowest-energy mean-field configuration. 
We find numerically that the squared radius of the axial PGCM state is closely approximated by the probability-weighted average of the individually projected squared radii. 
Since $\sum_\beta g^{I=0}(\beta)=1$, the shift relative to the mean-field result can therefore be written approximately as
\begin{align}
    [R_{\mathrm{c},\mathrm{PGCM}}^{I=0}]^2-[R_\mathrm{c}^{\mathrm{RDFT}}]^2 \simeq \sum_\beta g^{I=0}(\beta)[\Delta R_c^2(\beta)].
\end{align}

For $^{112}$Ru, the collective distribution peaks near $\beta=-0.2$, with a smaller component near $\beta=0.2$. 
The squared-radius differences near these peaks are negative, whereas configurations with large positive differences, particularly at larger $|\beta|$, carry little collective weight.
Negative contributions consequently dominate the weighted sum, explaining why the axial PGCM radius falls below the mean-field value. 
The same diagnostic can be applied to $^{106,108,110}$Ru.

%------------------------------------------------------------------
\begin{figure}[!htbp]
  \centering
  \includegraphics[width=0.5\textwidth]{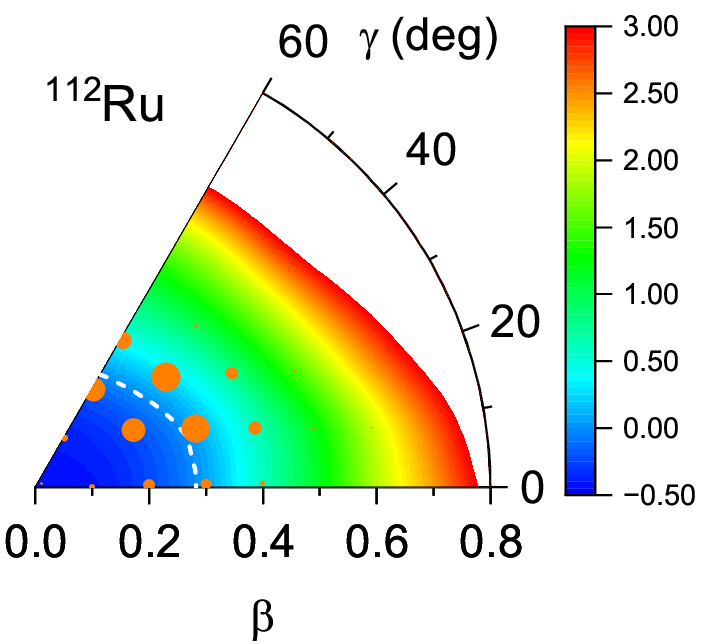}
  \caption{(Color online) Difference in squared charge radii, $\Delta R_\mathrm{c}^2(\beta,\gamma) \equiv [R_\mathrm{c}^{I=0}(\beta,\gamma)]^2-[R_c^{\mathrm{RDFT}}]^2$, for $^{112}$Ru (color map, in $\mathrm{fm}^2$), overlaid with the collective probability distribution $g^{I=0}(\beta,\gamma)$ from the full $(\beta,\gamma)$ PGCM calculation (orange circles). 
  Here, $R_\mathrm{c}^{I=0}(\beta,\gamma)$ is the radius of an individual intrinsic configuration after $I=0$ projection, without shape mixing, and $R_c^{\mathrm{RDFT}}$ is the radius of the lowest-energy mean-field configuration. 
  The size of each circle indicates the probability weight at that deformation point. 
  The white dashed line marks the $\Delta R_c^2=0$ contour, separating regions where the projected radius is smaller or larger than the mean-field reference.
  }
  \label{Fig:Triaxial}
\end{figure}
%------------------------------------------------------------------------------

To understand why the full PGCM calculation yields a larger charge radius than the mean-field calculation, we examine $^{112}$Ru in Fig.~\ref{Fig:Triaxial}. 
As in the axial case, we find numerically that the shift in the squared radius is well approximated by
\begin{align}
[R_{\mathrm{c},\mathrm{PGCM}}^{I=0}]^2-[R_\mathrm{c}^{\mathrm{RDFT}}]^2
\simeq
\sum_{\beta,\gamma} g^{I=0}(\beta,\gamma) [\Delta R_c^2(\beta,\gamma)],
\end{align}
where $\Delta R_\mathrm{c}^2(\beta,\gamma)$ is shown by the color map and $g^{I=0}(\beta,\gamma)$ by the orange circles. 
The white dashed zero contour separates configurations whose individually projected squared radii are smaller or larger than the mean-field reference.

The collective distribution extends into both regions. 
Configurations with $\Delta R_\mathrm{c}^2<0$ contribute $-0.0662~\mathrm{fm}^2$ to the weighted sum, whereas those with $\Delta R_\mathrm{c}^2>0$ contribute $+0.1868~\mathrm{fm}^2$. 
Their net contribution is therefore $+0.1206~\mathrm{fm}^2$: the positive weighted contributions outweigh the negative ones. 
This accounts for the increase in the full PGCM charge radius relative to the mean-field result and connects that increase to the distribution of the correlated ground state over the $(\beta,\gamma)$ plane.

 \textit{Summary}---We have investigated the charge radii of the even-even neutron-rich $^{106-114}$Ru isotopes using relativistic density-functional theory combined with three-dimensional angular-momentum projection and the generator-coordinate method. 
 By successively allowing static triaxial deformation, projected axial shape mixing, and shape mixing that includes projected triaxial configurations, we trace how the predicted radii change at each level. 
 At the mean-field level, static triaxiality increases the radii of $^{106,108}$Ru and improves their agreement with experiment, but does not reproduce the measured trend across the chain. 
 Projection and shape mixing restricted to axial configurations lower the radii of $^{106-112}$Ru relative to the corresponding mean-field results, widening the discrepancy with experiment. 
 Including projected triaxial configurations, by contrast, substantially raises the radii and brings them into close agreement with the measurements, without adjusting parameters to the Ru charge-radius data. 
 A deformation-resolved analysis links this enhancement to the interplay between collective weights and the radii of individually projected configurations: positive contributions to the squared-radius shift outweigh negative ones in the probability-weighted approximation. 
 These results demonstrate that quantum shape dynamics can substantially affect charge radii. 
 Precision charge-radius measurements, including those obtained by collinear laser spectroscopy, therefore provide benchmarks for correlated descriptions of nuclear shape, although they cannot alone uniquely determine the underlying intrinsic deformation.

\begin{acknowledgments}
This work has been supported in part by the National Natural Science Foundation of China under Grant No. 12105004, the Beijing Natural Science Foundation under Grant No. 1242019, the State Key Laboratory of Dark Matter Physics, and the Fundamental Research Funds for the Central Universities.
\end{acknowledgments}

\section*{DATA AVAILABILITY}
The data that support the findings of this article are not publicly available. 
The data are available from the authors upon reasonable request.

%\bibliography{reference}

\begin{thebibliography}{38}%
\makeatletter
\providecommand \@ifxundefined [1]{%
 \@ifx{#1\undefined}
}%
\providecommand \@ifnum [1]{%
 \ifnum #1\expandafter \@firstoftwo
 \else \expandafter \@secondoftwo
 \fi
}%
\providecommand \@ifx [1]{%
 \ifx #1\expandafter \@firstoftwo
 \else \expandafter \@secondoftwo
 \fi
}%
\providecommand \natexlab [1]{#1}%
\providecommand \enquote  [1]{``#1''}%
\providecommand \bibnamefont  [1]{#1}%
\providecommand \bibfnamefont [1]{#1}%
\providecommand \citenamefont [1]{#1}%
\providecommand \href@noop [0]{\@secondoftwo}%
\providecommand \href [0]{\begingroup \@sanitize@url \@href}%
\providecommand \@href[1]{\@@startlink{#1}\@@href}%
\providecommand \@@href[1]{\endgroup#1\@@endlink}%
\providecommand \@sanitize@url [0]{\catcode `\\12\catcode `\$12\catcode
  `\&12\catcode `\#12\catcode `\^12\catcode `\_12\catcode `\%12\relax}%
\providecommand \@@startlink[1]{}%
\providecommand \@@endlink[0]{}%
\providecommand \url  [0]{\begingroup\@sanitize@url \@url }%
\providecommand \@url [1]{\endgroup\@href {#1}{\urlprefix }}%
\providecommand \urlprefix  [0]{URL }%
\providecommand \Eprint [0]{\href }%
\providecommand \doibase [0]{https://doi.org/}%
\providecommand \selectlanguage [0]{\@gobble}%
\providecommand \bibinfo  [0]{\@secondoftwo}%
\providecommand \bibfield  [0]{\@secondoftwo}%
\providecommand \translation [1]{[#1]}%
\providecommand \BibitemOpen [0]{}%
\providecommand \bibitemStop [0]{}%
\providecommand \bibitemNoStop [0]{.\EOS\space}%
\providecommand \EOS [0]{\spacefactor3000\relax}%
\providecommand \BibitemShut  [1]{\csname bibitem#1\endcsname}%
\let\auto@bib@innerbib\@empty
%</preamble>
\bibitem [{\citenamefont {Gorges}\ \emph {et~al.}(2019)\citenamefont {Gorges}
  \emph {et~al.}}]{Gorges:2019wzy}%
  \BibitemOpen
  \bibfield  {author} {\bibinfo {author} {\bibfnamefont {C.}~\bibnamefont
  {Gorges}} \emph {et~al.},\ }\bibfield  {title} {\bibinfo {title} {{Laser
  spectroscopy of neutron-rich tin isotopes: A discontinuity in charge radii
  across the $N=82$ shell closure}},\ }\href
  {https://doi.org/10.1103/PhysRevLett.122.192502} {\bibfield  {journal}
  {\bibinfo  {journal} {Phys. Rev. Lett.}\ }\textbf {\bibinfo {volume} {122}},\
  \bibinfo {pages} {192502} (\bibinfo {year} {2019})}\BibitemShut {NoStop}%
\bibitem [{\citenamefont {Marsh}\ \emph {et~al.}(2018)\citenamefont {Marsh}
  \emph {et~al.}}]{Marsh:2018wxs}%
  \BibitemOpen
  \bibfield  {author} {\bibinfo {author} {\bibfnamefont {B.~A.}\ \bibnamefont
  {Marsh}} \emph {et~al.},\ }\bibfield  {title} {\bibinfo {title}
  {{Characterization of the shape-staggering effect in mercury nuclei}},\
  }\href {https://doi.org/10.1038/s41567-018-0292-8} {\bibfield  {journal}
  {\bibinfo  {journal} {Nature Phys.}\ }\textbf {\bibinfo {volume} {14}},\
  \bibinfo {pages} {1163} (\bibinfo {year} {2018})},\ \bibinfo {note}
  {[Erratum: Nature Phys. 18, 219 (2022)]}\BibitemShut {NoStop}%
\bibitem [{\citenamefont {Barzakh}\ \emph {et~al.}(2021)\citenamefont {Barzakh}
  \emph {et~al.}}]{Barzakh:2021gfl}%
  \BibitemOpen
  \bibfield  {author} {\bibinfo {author} {\bibfnamefont {A.}~\bibnamefont
  {Barzakh}} \emph {et~al.},\ }\bibfield  {title} {\bibinfo {title} {{Large
  Shape Staggering in Neutron-Deficient Bi Isotopes}},\ }\href
  {https://doi.org/10.1103/PhysRevLett.127.192501} {\bibfield  {journal}
  {\bibinfo  {journal} {Phys. Rev. Lett.}\ }\textbf {\bibinfo {volume} {127}},\
  \bibinfo {pages} {192501} (\bibinfo {year} {2021})}\BibitemShut {NoStop}%
\bibitem [{\citenamefont {Yang}\ \emph {et~al.}(2016)\citenamefont {Yang} \emph
  {et~al.}}]{Yang:2016tyv}%
  \BibitemOpen
  \bibfield  {author} {\bibinfo {author} {\bibfnamefont {X.~F.}\ \bibnamefont
  {Yang}} \emph {et~al.},\ }\bibfield  {title} {\bibinfo {title} {{Isomer shift
  and magnetic moment of the long-lived 1/2$^{+}$ isomer in
  $^{79}_{30}$Zn$_{49}$: signature of shape coexistence near $^{78}$Ni}},\
  }\href {https://doi.org/10.1103/PhysRevLett.116.219901} {\bibfield  {journal}
  {\bibinfo  {journal} {Phys. Rev. Lett.}\ }\textbf {\bibinfo {volume} {116}},\
  \bibinfo {pages} {182502} (\bibinfo {year} {2016})},\ \bibinfo {note}
  {[Addendum: Phys.Rev.Lett. 116, 219901 (2016)]},\ \Eprint
  {https://arxiv.org/abs/1604.03316} {arXiv:1604.03316 [nucl-ex]} \BibitemShut
  {NoStop}%
\bibitem [{\citenamefont {Hagen}\ \emph {et~al.}(2015)\citenamefont {Hagen}
  \emph {et~al.}}]{Hagen:2015yea}%
  \BibitemOpen
  \bibfield  {author} {\bibinfo {author} {\bibfnamefont {G.}~\bibnamefont
  {Hagen}} \emph {et~al.},\ }\bibfield  {title} {\bibinfo {title} {{Neutron and
  weak-charge distributions of the $^{48}$Ca nucleus}},\ }\href
  {https://doi.org/10.1038/nphys3529} {\bibfield  {journal} {\bibinfo
  {journal} {Nature Phys.}\ }\textbf {\bibinfo {volume} {12}},\ \bibinfo
  {pages} {186} (\bibinfo {year} {2015})},\ \Eprint
  {https://arxiv.org/abs/1509.07169} {arXiv:1509.07169 [nucl-th]} \BibitemShut
  {NoStop}%
\bibitem [{\citenamefont {Kreim}\ \emph {et~al.}(2014)\citenamefont {Kreim}
  \emph {et~al.}}]{Kreim:2013uqb}%
  \BibitemOpen
  \bibfield  {author} {\bibinfo {author} {\bibfnamefont {K.}~\bibnamefont
  {Kreim}} \emph {et~al.},\ }\bibfield  {title} {\bibinfo {title} {{Nuclear
  charge radii of potassium isotopes beyond $N=28$}},\ }\href
  {https://doi.org/10.1016/j.physletb.2014.02.012} {\bibfield  {journal}
  {\bibinfo  {journal} {Phys. Lett. B}\ }\textbf {\bibinfo {volume} {731}},\
  \bibinfo {pages} {97} (\bibinfo {year} {2014})},\ \Eprint
  {https://arxiv.org/abs/1310.5171} {arXiv:1310.5171 [nucl-ex]} \BibitemShut
  {NoStop}%
\bibitem [{\citenamefont {Bohr}\ and\ \citenamefont
  {Mottelson}(1975)}]{BohrMottelson1975}%
  \BibitemOpen
  \bibfield  {author} {\bibinfo {author} {\bibfnamefont {A.}~\bibnamefont
  {Bohr}}\ and\ \bibinfo {author} {\bibfnamefont {B.~R.}\ \bibnamefont
  {Mottelson}},\ }\href@noop {} {\emph {\bibinfo {title} {Nuclear Structure,
  Volume II: Nuclear Deformations}}}\ (\bibinfo  {publisher} {W. A. Benjamin,
  Inc.},\ \bibinfo {address} {Reading, Massachusetts},\ \bibinfo {year}
  {1975})\ p.\ \bibinfo {pages} {748}\BibitemShut {NoStop}%
\bibitem [{\citenamefont {Niksic}\ \emph {et~al.}(2011)\citenamefont {Niksic},
  \citenamefont {Vretenar},\ and\ \citenamefont {Ring}}]{Niksic:2011sg}%
  \BibitemOpen
  \bibfield  {author} {\bibinfo {author} {\bibfnamefont {T.}~\bibnamefont
  {Niksic}}, \bibinfo {author} {\bibfnamefont {D.}~\bibnamefont {Vretenar}},\
  and\ \bibinfo {author} {\bibfnamefont {P.}~\bibnamefont {Ring}},\ }\bibfield
  {title} {\bibinfo {title} {{Relativistic Nuclear Energy Density Functionals:
  Mean-Field and Beyond}},\ }\href {https://doi.org/10.1016/j.ppnp.2011.01.055}
  {\bibfield  {journal} {\bibinfo  {journal} {Prog. Part. Nucl. Phys.}\
  }\textbf {\bibinfo {volume} {66}},\ \bibinfo {pages} {519} (\bibinfo {year}
  {2011})},\ \Eprint {https://arxiv.org/abs/1102.4193} {arXiv:1102.4193
  [nucl-th]} \BibitemShut {NoStop}%
\bibitem [{\citenamefont {Maass}\ \emph {et~al.}(2025)\citenamefont {Maass}
  \emph {et~al.}}]{Maass:2025gnq}%
  \BibitemOpen
  \bibfield  {author} {\bibinfo {author} {\bibfnamefont {B.}~\bibnamefont
  {Maass}} \emph {et~al.},\ }\bibfield  {title} {\bibinfo {title}
  {{Fingerprints of Triaxiality in the Charge Radii of Neutron-Rich
  Ruthenium}},\ }\href {https://doi.org/10.1103/81h5-wjkd} {\bibfield
  {journal} {\bibinfo  {journal} {Phys. Rev. Lett.}\ }\textbf {\bibinfo
  {volume} {135}},\ \bibinfo {pages} {202501} (\bibinfo {year} {2025})},\
  \Eprint {https://arxiv.org/abs/2503.07841} {arXiv:2503.07841 [nucl-ex]}
  \BibitemShut {NoStop}%
\bibitem [{\citenamefont {Yang}\ \emph {et~al.}(2021)\citenamefont {Yang},
  \citenamefont {Wang}, \citenamefont {Zhao},\ and\ \citenamefont
  {Li}}]{Yang:2021wcd}%
  \BibitemOpen
  \bibfield  {author} {\bibinfo {author} {\bibfnamefont {Y.~L.}\ \bibnamefont
  {Yang}}, \bibinfo {author} {\bibfnamefont {Y.~K.}\ \bibnamefont {Wang}},
  \bibinfo {author} {\bibfnamefont {P.~W.}\ \bibnamefont {Zhao}},\ and\
  \bibinfo {author} {\bibfnamefont {Z.~P.}\ \bibnamefont {Li}},\ }\bibfield
  {title} {\bibinfo {title} {{Nuclear landscape in a mapped collective
  Hamiltonian from covariant density functional theory}},\ }\href
  {https://doi.org/10.1103/PhysRevC.104.054312} {\bibfield  {journal} {\bibinfo
   {journal} {Phys. Rev. C}\ }\textbf {\bibinfo {volume} {104}},\ \bibinfo
  {pages} {054312} (\bibinfo {year} {2021})},\ \Eprint
  {https://arxiv.org/abs/2108.13057} {arXiv:2108.13057 [nucl-th]} \BibitemShut
  {NoStop}%
\bibitem [{\citenamefont {Yang}\ \emph {et~al.}(2023)\citenamefont {Yang},
  \citenamefont {Zhao},\ and\ \citenamefont {Li}}]{Yang:2023dql}%
  \BibitemOpen
  \bibfield  {author} {\bibinfo {author} {\bibfnamefont {Y.~L.}\ \bibnamefont
  {Yang}}, \bibinfo {author} {\bibfnamefont {P.~W.}\ \bibnamefont {Zhao}},\
  and\ \bibinfo {author} {\bibfnamefont {Z.~P.}\ \bibnamefont {Li}},\
  }\bibfield  {title} {\bibinfo {title} {{Shape and multiple shape coexistence
  of nuclei within covariant density functional theory}},\ }\href
  {https://doi.org/10.1103/PhysRevC.107.024308} {\bibfield  {journal} {\bibinfo
   {journal} {Phys. Rev. C}\ }\textbf {\bibinfo {volume} {107}},\ \bibinfo
  {pages} {024308} (\bibinfo {year} {2023})},\ \Eprint
  {https://arxiv.org/abs/2302.07417} {arXiv:2302.07417 [nucl-th]} \BibitemShut
  {NoStop}%
\bibitem [{\citenamefont {Egido}\ and\ \citenamefont
  {Jungclaus}(2020)}]{Egido:2020bql}%
  \BibitemOpen
  \bibfield  {author} {\bibinfo {author} {\bibfnamefont {J.~L.}\ \bibnamefont
  {Egido}}\ and\ \bibinfo {author} {\bibfnamefont {A.}~\bibnamefont
  {Jungclaus}},\ }\bibfield  {title} {\bibinfo {title} {{Predominance of
  Triaxial Shapes in Transitional Super-Heavy Nuclei: Ground-State Deformation
  and Shape Coexistence along the Flerovium ( Z=114 ) Chain of Isotopes}},\
  }\href {https://doi.org/10.1103/PhysRevLett.125.192504} {\bibfield  {journal}
  {\bibinfo  {journal} {Phys. Rev. Lett.}\ }\textbf {\bibinfo {volume} {125}},\
  \bibinfo {pages} {192504} (\bibinfo {year} {2020})}\BibitemShut {NoStop}%
\bibitem [{\citenamefont {Rodr{\'\i}guez}\ and\ \citenamefont
  {Egido}(2011)}]{Rodriguez:2011hb}%
  \BibitemOpen
  \bibfield  {author} {\bibinfo {author} {\bibfnamefont {T.~R.}\ \bibnamefont
  {Rodr{\'\i}guez}}\ and\ \bibinfo {author} {\bibfnamefont {J.~L.}\
  \bibnamefont {Egido}},\ }\bibfield  {title} {\bibinfo {title} {{Multiple
  shape coexistence in the nucleus $^{80}$Zr}},\ }\href
  {https://doi.org/10.1016/j.physletb.2011.10.003} {\bibfield  {journal}
  {\bibinfo  {journal} {Phys. Lett. B}\ }\textbf {\bibinfo {volume} {705}},\
  \bibinfo {pages} {255} (\bibinfo {year} {2011})},\ \Eprint
  {https://arxiv.org/abs/1109.6511} {arXiv:1109.6511 [nucl-th]} \BibitemShut
  {NoStop}%
\bibitem [{\citenamefont {Nomura}\ \emph {et~al.}(2017)\citenamefont {Nomura},
  \citenamefont {Nik{\v{s}}i{\'c}},\ and\ \citenamefont
  {Vretenar}}]{Nomura:2017ilh}%
  \BibitemOpen
  \bibfield  {author} {\bibinfo {author} {\bibfnamefont {K.}~\bibnamefont
  {Nomura}}, \bibinfo {author} {\bibfnamefont {T.}~\bibnamefont
  {Nik{\v{s}}i{\'c}}},\ and\ \bibinfo {author} {\bibfnamefont {D.}~\bibnamefont
  {Vretenar}},\ }\bibfield  {title} {\bibinfo {title} {{Shape-phase transitions
  in odd-mass {\ensuremath{\gamma}} -soft nuclei with mass
  A{\ensuremath{\approx}}130}},\ }\href
  {https://doi.org/10.1103/PhysRevC.96.014304} {\bibfield  {journal} {\bibinfo
  {journal} {Phys. Rev. C}\ }\textbf {\bibinfo {volume} {96}},\ \bibinfo
  {pages} {014304} (\bibinfo {year} {2017})},\ \Eprint
  {https://arxiv.org/abs/1704.07101} {arXiv:1704.07101 [nucl-th]} \BibitemShut
  {NoStop}%
\bibitem [{\citenamefont {Li}\ \emph {et~al.}(2010)\citenamefont {Li},
  \citenamefont {Niksic}, \citenamefont {Vretenar},\ and\ \citenamefont
  {Meng}}]{Li:2010qu}%
  \BibitemOpen
  \bibfield  {author} {\bibinfo {author} {\bibfnamefont {Z.~P.}\ \bibnamefont
  {Li}}, \bibinfo {author} {\bibfnamefont {T.}~\bibnamefont {Niksic}}, \bibinfo
  {author} {\bibfnamefont {D.}~\bibnamefont {Vretenar}},\ and\ \bibinfo
  {author} {\bibfnamefont {J.}~\bibnamefont {Meng}},\ }\bibfield  {title}
  {\bibinfo {title} {{Microscopic Description of Spherical to Gamma-Soft Shape
  Transitions in Ba and Xe Nuclei}},\ }\href
  {https://doi.org/10.1103/PhysRevC.81.034316} {\bibfield  {journal} {\bibinfo
  {journal} {Phys. Rev. C}\ }\textbf {\bibinfo {volume} {81}},\ \bibinfo
  {pages} {034316} (\bibinfo {year} {2010})},\ \Eprint
  {https://arxiv.org/abs/1003.4109} {arXiv:1003.4109 [nucl-th]} \BibitemShut
  {NoStop}%
\bibitem [{\citenamefont {Togashi}\ \emph {et~al.}(2016)\citenamefont
  {Togashi}, \citenamefont {Tsunoda}, \citenamefont {Otsuka},\ and\
  \citenamefont {Shimizu}}]{Togashi:2016yzs}%
  \BibitemOpen
  \bibfield  {author} {\bibinfo {author} {\bibfnamefont {T.}~\bibnamefont
  {Togashi}}, \bibinfo {author} {\bibfnamefont {Y.}~\bibnamefont {Tsunoda}},
  \bibinfo {author} {\bibfnamefont {T.}~\bibnamefont {Otsuka}},\ and\ \bibinfo
  {author} {\bibfnamefont {N.}~\bibnamefont {Shimizu}},\ }\bibfield  {title}
  {\bibinfo {title} {{Quantum Phase Transition in the Shape of Zr isotopes}},\
  }\href {https://doi.org/10.1103/PhysRevLett.117.172502} {\bibfield  {journal}
  {\bibinfo  {journal} {Phys. Rev. Lett.}\ }\textbf {\bibinfo {volume} {117}},\
  \bibinfo {pages} {172502} (\bibinfo {year} {2016})},\ \Eprint
  {https://arxiv.org/abs/1606.09056} {arXiv:1606.09056 [nucl-th]} \BibitemShut
  {NoStop}%
\bibitem [{\citenamefont {Chen}\ and\ \citenamefont
  {Egido}(2017)}]{Chen:2017vtf}%
  \BibitemOpen
  \bibfield  {author} {\bibinfo {author} {\bibfnamefont {F.-Q.}\ \bibnamefont
  {Chen}}\ and\ \bibinfo {author} {\bibfnamefont {J.~L.}\ \bibnamefont
  {Egido}},\ }\bibfield  {title} {\bibinfo {title} {{Triaxial shape
  fluctuations and quasiparticle excitations in heavy nuclei}},\ }\href
  {https://doi.org/10.1103/PhysRevC.95.024307} {\bibfield  {journal} {\bibinfo
  {journal} {Phys. Rev. C}\ }\textbf {\bibinfo {volume} {95}},\ \bibinfo
  {pages} {024307} (\bibinfo {year} {2017})},\ \Eprint
  {https://arxiv.org/abs/1701.03650} {arXiv:1701.03650 [nucl-th]} \BibitemShut
  {NoStop}%
\bibitem [{\citenamefont {Rodriguez}\ and\ \citenamefont
  {Egido}(2010)}]{Rodriguez:2010by}%
  \BibitemOpen
  \bibfield  {author} {\bibinfo {author} {\bibfnamefont {T.~R.}\ \bibnamefont
  {Rodriguez}}\ and\ \bibinfo {author} {\bibfnamefont {J.~L.}\ \bibnamefont
  {Egido}},\ }\bibfield  {title} {\bibinfo {title} {{Triaxial Angular Momentum
  Projection and Configuration Mixing calculations with the Gogny force}},\
  }\href {https://doi.org/10.1103/PhysRevC.81.064323} {\bibfield  {journal}
  {\bibinfo  {journal} {Phys. Rev. C}\ }\textbf {\bibinfo {volume} {81}},\
  \bibinfo {pages} {064323} (\bibinfo {year} {2010})},\ \Eprint
  {https://arxiv.org/abs/1004.2877} {arXiv:1004.2877 [nucl-th]} \BibitemShut
  {NoStop}%
\bibitem [{\citenamefont {Xiang}\ \emph {et~al.}(2024)\citenamefont {Xiang},
  \citenamefont {Li}, \citenamefont {Nik{\v{s}}i{\'c}}, \citenamefont
  {Vretenar}, \citenamefont {Long},\ and\ \citenamefont {Wu}}]{Xiang:2023eer}%
  \BibitemOpen
  \bibfield  {author} {\bibinfo {author} {\bibfnamefont {J.}~\bibnamefont
  {Xiang}}, \bibinfo {author} {\bibfnamefont {Z.~P.}\ \bibnamefont {Li}},
  \bibinfo {author} {\bibfnamefont {T.}~\bibnamefont {Nik{\v{s}}i{\'c}}},
  \bibinfo {author} {\bibfnamefont {D.}~\bibnamefont {Vretenar}}, \bibinfo
  {author} {\bibfnamefont {W.~H.}\ \bibnamefont {Long}},\ and\ \bibinfo
  {author} {\bibfnamefont {X.~Y.}\ \bibnamefont {Wu}},\ }\bibfield  {title}
  {\bibinfo {title} {{Coupling of shape and pairing vibrations in a collective
  Hamiltonian based on nuclear energy density functionals. II. Low-energy
  excitation spectra of triaxial nuclei}},\ }\href
  {https://doi.org/10.1103/PhysRevC.109.044319} {\bibfield  {journal} {\bibinfo
   {journal} {Phys. Rev. C}\ }\textbf {\bibinfo {volume} {109}},\ \bibinfo
  {pages} {044319} (\bibinfo {year} {2024})},\ \Eprint
  {https://arxiv.org/abs/2312.01791} {arXiv:2312.01791 [nucl-th]} \BibitemShut
  {NoStop}%
\bibitem [{\citenamefont {Jiao}\ \emph {et~al.}(2017)\citenamefont {Jiao},
  \citenamefont {Engel},\ and\ \citenamefont {Holt}}]{Jiao:2017opc}%
  \BibitemOpen
  \bibfield  {author} {\bibinfo {author} {\bibfnamefont {C.~F.}\ \bibnamefont
  {Jiao}}, \bibinfo {author} {\bibfnamefont {J.}~\bibnamefont {Engel}},\ and\
  \bibinfo {author} {\bibfnamefont {J.~D.}\ \bibnamefont {Holt}},\ }\bibfield
  {title} {\bibinfo {title} {{Neutrinoless double-beta decay matrix elements in
  large shell-model spaces with the generator-coordinate method}},\ }\href
  {https://doi.org/10.1103/PhysRevC.96.054310} {\bibfield  {journal} {\bibinfo
  {journal} {Phys. Rev. C}\ }\textbf {\bibinfo {volume} {96}},\ \bibinfo
  {pages} {054310} (\bibinfo {year} {2017})},\ \Eprint
  {https://arxiv.org/abs/1707.03940} {arXiv:1707.03940 [nucl-th]} \BibitemShut
  {NoStop}%
\bibitem [{\citenamefont {Wang}\ \emph {et~al.}(2024)\citenamefont {Wang},
  \citenamefont {Zhao},\ and\ \citenamefont {Meng}}]{Wang:2023hkc}%
  \BibitemOpen
  \bibfield  {author} {\bibinfo {author} {\bibfnamefont {Y.}~\bibnamefont
  {Wang}}, \bibinfo {author} {\bibfnamefont {P.}~\bibnamefont {Zhao}},\ and\
  \bibinfo {author} {\bibfnamefont {J.}~\bibnamefont {Meng}},\ }\bibfield
  {title} {\bibinfo {title} {{Relativistic configuration-interaction density
  functional theory: Nonaxial effects on nuclear
  {\ensuremath{\beta}}{\ensuremath{\beta}} decay}},\ }\href
  {https://doi.org/10.1016/j.scib.2024.04.071} {\bibfield  {journal} {\bibinfo
  {journal} {Sci. Bull.}\ }\textbf {\bibinfo {volume} {69}},\ \bibinfo {pages}
  {2017} (\bibinfo {year} {2024})},\ \Eprint {https://arxiv.org/abs/2304.12009}
  {arXiv:2304.12009 [nucl-th]} \BibitemShut {NoStop}%
\bibitem [{\citenamefont {Libert}\ \emph {et~al.}(2007)\citenamefont {Libert},
  \citenamefont {Roussi{\`e}re},\ and\ \citenamefont
  {Sauvage}}]{Libert:2007mrz}%
  \BibitemOpen
  \bibfield  {author} {\bibinfo {author} {\bibfnamefont {J.}~\bibnamefont
  {Libert}}, \bibinfo {author} {\bibfnamefont {B.}~\bibnamefont
  {Roussi{\`e}re}},\ and\ \bibinfo {author} {\bibfnamefont {J.}~\bibnamefont
  {Sauvage}},\ }\bibfield  {title} {\bibinfo {title} {{Signs of dynamical
  effects for Cd, Sn, Te, Xe, Ba and Sm nuclear charge radii}},\ }\href
  {https://doi.org/10.1016/j.nuclphysa.2007.01.089} {\bibfield  {journal}
  {\bibinfo  {journal} {Nucl. Phys. A}\ }\textbf {\bibinfo {volume} {786}},\
  \bibinfo {pages} {47} (\bibinfo {year} {2007})}\BibitemShut {NoStop}%
\bibitem [{Nuc()}]{NuclearMap}%
  \BibitemOpen
  \href@noop {} {}\bibinfo {note}
  {\url{http://nuclearmap.jcnp.org/}}\BibitemShut {NoStop}%
\bibitem [{\citenamefont {Esmaylzadeh}\ \emph {et~al.}(2022)\citenamefont
  {Esmaylzadeh} \emph {et~al.}}]{Esmaylzadeh:2022xgi}%
  \BibitemOpen
  \bibfield  {author} {\bibinfo {author} {\bibfnamefont {A.}~\bibnamefont
  {Esmaylzadeh}} \emph {et~al.},\ }\bibfield  {title} {\bibinfo {title}
  {{Investigation of {\ensuremath{\gamma}} softness: Lifetime measurements in
  Ru104,106}},\ }\href {https://doi.org/10.1103/PhysRevC.106.064323} {\bibfield
   {journal} {\bibinfo  {journal} {Phys. Rev. C}\ }\textbf {\bibinfo {volume}
  {106}},\ \bibinfo {pages} {064323} (\bibinfo {year} {2022})}\BibitemShut
  {NoStop}%
\bibitem [{\citenamefont {Garrett}\ \emph {et~al.}(2020)\citenamefont {Garrett}
  \emph {et~al.}}]{Garrett:2020ded}%
  \BibitemOpen
  \bibfield  {author} {\bibinfo {author} {\bibfnamefont {P.~E.}\ \bibnamefont
  {Garrett}} \emph {et~al.},\ }\bibfield  {title} {\bibinfo {title}
  {{Observation of the 02+ and {\ensuremath{\gamma}} bands in 98 Ru, and shape
  coexistence in the Ru isotopes}},\ }\href
  {https://doi.org/10.1016/j.physletb.2020.135762} {\bibfield  {journal}
  {\bibinfo  {journal} {Phys. Lett. B}\ }\textbf {\bibinfo {volume} {809}},\
  \bibinfo {pages} {135762} (\bibinfo {year} {2020})}\BibitemShut {NoStop}%
\bibitem [{\citenamefont {Nomura}\ \emph {et~al.}(2016)\citenamefont {Nomura},
  \citenamefont {Rodr{\'\i}guez-Guzm{\'a}n},\ and\ \citenamefont
  {Robledo}}]{Nomura:2016nyh}%
  \BibitemOpen
  \bibfield  {author} {\bibinfo {author} {\bibfnamefont {K.}~\bibnamefont
  {Nomura}}, \bibinfo {author} {\bibfnamefont {R.}~\bibnamefont
  {Rodr{\'\i}guez-Guzm{\'a}n}},\ and\ \bibinfo {author} {\bibfnamefont {L.~M.}\
  \bibnamefont {Robledo}},\ }\bibfield  {title} {\bibinfo {title} {{Structural
  evolution in A{\ensuremath{\approx}}100 nuclei within the mapped interacting
  boson model based on the Gogny energy density functional}},\ }\href
  {https://doi.org/10.1103/PhysRevC.94.044314} {\bibfield  {journal} {\bibinfo
  {journal} {Phys. Rev. C}\ }\textbf {\bibinfo {volume} {94}},\ \bibinfo
  {pages} {044314} (\bibinfo {year} {2016})},\ \Eprint
  {https://arxiv.org/abs/1609.09614} {arXiv:1609.09614 [nucl-th]} \BibitemShut
  {NoStop}%
\bibitem [{\citenamefont {Ring}\ and\ \citenamefont
  {Schuck}(2004)}]{Ring:2004}%
  \BibitemOpen
  \bibfield  {author} {\bibinfo {author} {\bibfnamefont {P.}~\bibnamefont
  {Ring}}\ and\ \bibinfo {author} {\bibfnamefont {P.}~\bibnamefont {Schuck}},\
  }\href@noop {} {\emph {\bibinfo {title} {The Nuclear Many-Body Problem}}}\
  (\bibinfo  {publisher} {Springer Science \& Business Media},\ \bibinfo
  {address} {New York},\ \bibinfo {year} {2004})\BibitemShut {NoStop}%
\bibitem [{\citenamefont {Meng}(2016)}]{Meng2016}%
  \BibitemOpen
  \bibinfo {editor} {\bibfnamefont {J.}~\bibnamefont {Meng}},\ ed.,\ \href
  {https://doi.org/10.1142/9872} {\emph {\bibinfo {title} {Relativistic Density
  Functional for Nuclear Structure}}},\ \bibinfo {series} {International Review
  of Nuclear Physics}, Vol.~\bibinfo {volume} {10}\ (\bibinfo  {publisher}
  {World Scientific},\ \bibinfo {address} {Singapore},\ \bibinfo {year}
  {2016})\BibitemShut {NoStop}%
\bibitem [{\citenamefont {Yao}\ \emph {et~al.}(2009)\citenamefont {Yao},
  \citenamefont {Meng}, \citenamefont {Ring},\ and\ \citenamefont
  {Pena~Arteaga}}]{Yao:2009ry}%
  \BibitemOpen
  \bibfield  {author} {\bibinfo {author} {\bibfnamefont {J.~M.}\ \bibnamefont
  {Yao}}, \bibinfo {author} {\bibfnamefont {J.}~\bibnamefont {Meng}}, \bibinfo
  {author} {\bibfnamefont {P.}~\bibnamefont {Ring}},\ and\ \bibinfo {author}
  {\bibfnamefont {D.}~\bibnamefont {Pena~Arteaga}},\ }\bibfield  {title}
  {\bibinfo {title} {{Three-dimensional angular momentum projection in
  relativistic mean-field theory}},\ }\href
  {https://doi.org/10.1103/PhysRevC.79.044312} {\bibfield  {journal} {\bibinfo
  {journal} {Phys. Rev. C}\ }\textbf {\bibinfo {volume} {79}},\ \bibinfo
  {pages} {044312} (\bibinfo {year} {2009})},\ \Eprint
  {https://arxiv.org/abs/0903.5027} {arXiv:0903.5027 [nucl-th]} \BibitemShut
  {NoStop}%
\bibitem [{\citenamefont {Egido}(2016)}]{Egido:2016bdz}%
  \BibitemOpen
  \bibfield  {author} {\bibinfo {author} {\bibfnamefont {J.~L.}\ \bibnamefont
  {Egido}},\ }\bibfield  {title} {\bibinfo {title} {{State-of-the-art of beyond
  mean field theories with nuclear density functionals}},\ }\href
  {https://doi.org/10.1088/0031-8949/91/7/073003} {\bibfield  {journal}
  {\bibinfo  {journal} {Phys. Scripta}\ }\textbf {\bibinfo {volume} {91}},\
  \bibinfo {pages} {073003} (\bibinfo {year} {2016})},\ \Eprint
  {https://arxiv.org/abs/1606.00407} {arXiv:1606.00407 [nucl-th]} \BibitemShut
  {NoStop}%
\bibitem [{\citenamefont {Hu}\ \emph {et~al.}(2014)\citenamefont {Hu},
  \citenamefont {Gao},\ and\ \citenamefont {Chen}}]{Hu:2013dha}%
  \BibitemOpen
  \bibfield  {author} {\bibinfo {author} {\bibfnamefont {Q.-L.}\ \bibnamefont
  {Hu}}, \bibinfo {author} {\bibfnamefont {Z.-C.}\ \bibnamefont {Gao}},\ and\
  \bibinfo {author} {\bibfnamefont {Y.~S.}\ \bibnamefont {Chen}},\ }\bibfield
  {title} {\bibinfo {title} {{Matrix elements of one-body and two-body
  operators between arbitrary HFB multi-quasiparticle states}},\ }\href
  {https://doi.org/10.1016/j.physletb.2014.05.045} {\bibfield  {journal}
  {\bibinfo  {journal} {Phys. Lett. B}\ }\textbf {\bibinfo {volume} {734}},\
  \bibinfo {pages} {162} (\bibinfo {year} {2014})},\ \Eprint
  {https://arxiv.org/abs/1307.6905} {arXiv:1307.6905 [nucl-th]} \BibitemShut
  {NoStop}%
\bibitem [{\citenamefont {Carlsson}\ and\ \citenamefont
  {Rotureau}(2021)}]{Carlsson:2020fiv}%
  \BibitemOpen
  \bibfield  {author} {\bibinfo {author} {\bibfnamefont {B.~G.}\ \bibnamefont
  {Carlsson}}\ and\ \bibinfo {author} {\bibfnamefont {J.}~\bibnamefont
  {Rotureau}},\ }\bibfield  {title} {\bibinfo {title} {{New and Practical
  Formulation for Overlaps of Bogoliubov Vacua}},\ }\href
  {https://doi.org/10.1103/PhysRevLett.126.172501} {\bibfield  {journal}
  {\bibinfo  {journal} {Phys. Rev. Lett.}\ }\textbf {\bibinfo {volume} {126}},\
  \bibinfo {pages} {172501} (\bibinfo {year} {2021})},\ \Eprint
  {https://arxiv.org/abs/2010.08459} {arXiv:2010.08459 [nucl-th]} \BibitemShut
  {NoStop}%
\bibitem [{\citenamefont {Niksic}\ \emph {et~al.}(2014)\citenamefont {Niksic},
  \citenamefont {Paar}, \citenamefont {Vretenar},\ and\ \citenamefont
  {Ring}}]{Niksic:2014dra}%
  \BibitemOpen
  \bibfield  {author} {\bibinfo {author} {\bibfnamefont {T.}~\bibnamefont
  {Niksic}}, \bibinfo {author} {\bibfnamefont {N.}~\bibnamefont {Paar}},
  \bibinfo {author} {\bibfnamefont {D.}~\bibnamefont {Vretenar}},\ and\
  \bibinfo {author} {\bibfnamefont {P.}~\bibnamefont {Ring}},\ }\bibfield
  {title} {\bibinfo {title} {{DIRHB - A relativistic self-consistent mean-field
  framework for atomic nuclei}},\ }\href
  {https://doi.org/10.1016/j.cpc.2014.02.027} {\bibfield  {journal} {\bibinfo
  {journal} {Comput. Phys. Commun.}\ }\textbf {\bibinfo {volume} {185}},\
  \bibinfo {pages} {1808} (\bibinfo {year} {2014})},\ \Eprint
  {https://arxiv.org/abs/1403.4039} {arXiv:1403.4039 [nucl-th]} \BibitemShut
  {NoStop}%
\bibitem [{\citenamefont {Bender}\ \emph {et~al.}(2006)\citenamefont {Bender},
  \citenamefont {Bertsch},\ and\ \citenamefont {Heenen}}]{Bender:2005ri}%
  \BibitemOpen
  \bibfield  {author} {\bibinfo {author} {\bibfnamefont {M.}~\bibnamefont
  {Bender}}, \bibinfo {author} {\bibfnamefont {G.~F.}\ \bibnamefont
  {Bertsch}},\ and\ \bibinfo {author} {\bibfnamefont {P.~H.}\ \bibnamefont
  {Heenen}},\ }\bibfield  {title} {\bibinfo {title} {{Global study of
  quadrupole correlation effects}},\ }\href
  {https://doi.org/10.1103/PhysRevC.73.034322} {\bibfield  {journal} {\bibinfo
  {journal} {Phys. Rev. C}\ }\textbf {\bibinfo {volume} {73}},\ \bibinfo
  {pages} {034322} (\bibinfo {year} {2006})},\ \Eprint
  {https://arxiv.org/abs/nucl-th/0508052} {arXiv:nucl-th/0508052} \BibitemShut
  {NoStop}%
\bibitem [{\citenamefont {Zhao}\ \emph {et~al.}(2010)\citenamefont {Zhao},
  \citenamefont {Li}, \citenamefont {Yao},\ and\ \citenamefont
  {Meng}}]{Zhao:2010hi}%
  \BibitemOpen
  \bibfield  {author} {\bibinfo {author} {\bibfnamefont {P.~W.}\ \bibnamefont
  {Zhao}}, \bibinfo {author} {\bibfnamefont {Z.~P.}\ \bibnamefont {Li}},
  \bibinfo {author} {\bibfnamefont {J.~M.}\ \bibnamefont {Yao}},\ and\ \bibinfo
  {author} {\bibfnamefont {J.}~\bibnamefont {Meng}},\ }\bibfield  {title}
  {\bibinfo {title} {{New parametrization for the nuclear covariant energy
  density functional with point-coupling interaction}},\ }\href
  {https://doi.org/10.1103/PhysRevC.82.054319} {\bibfield  {journal} {\bibinfo
  {journal} {Phys. Rev. C}\ }\textbf {\bibinfo {volume} {82}},\ \bibinfo
  {pages} {054319} (\bibinfo {year} {2010})},\ \Eprint
  {https://arxiv.org/abs/1002.1789} {arXiv:1002.1789 [nucl-th]} \BibitemShut
  {NoStop}%
\bibitem [{\citenamefont {Tian}\ \emph {et~al.}(2009)\citenamefont {Tian},
  \citenamefont {Ma},\ and\ \citenamefont {Ring}}]{Tian:2009zzh}%
  \BibitemOpen
  \bibfield  {author} {\bibinfo {author} {\bibfnamefont {Y.}~\bibnamefont
  {Tian}}, \bibinfo {author} {\bibfnamefont {Z.~Y.}\ \bibnamefont {Ma}},\ and\
  \bibinfo {author} {\bibfnamefont {P.}~\bibnamefont {Ring}},\ }\bibfield
  {title} {\bibinfo {title} {{A finite range pairing force for density
  functional theory in superfluid nuclei}},\ }\href
  {https://doi.org/10.1016/j.physletb.2009.04.067} {\bibfield  {journal}
  {\bibinfo  {journal} {Phys. Lett. B}\ }\textbf {\bibinfo {volume} {676}},\
  \bibinfo {pages} {44} (\bibinfo {year} {2009})},\ \Eprint
  {https://arxiv.org/abs/0908.1844} {arXiv:0908.1844 [nucl-th]} \BibitemShut
  {NoStop}%
\bibitem [{\citenamefont {Bonche}\ \emph {et~al.}(1990)\citenamefont {Bonche},
  \citenamefont {Dobaczewski}, \citenamefont {Flocard}, \citenamefont
  {Heenen},\ and\ \citenamefont {Meyer}}]{Bonche:1990ghh}%
  \BibitemOpen
  \bibfield  {author} {\bibinfo {author} {\bibfnamefont {P.}~\bibnamefont
  {Bonche}}, \bibinfo {author} {\bibfnamefont {J.}~\bibnamefont {Dobaczewski}},
  \bibinfo {author} {\bibfnamefont {H.}~\bibnamefont {Flocard}}, \bibinfo
  {author} {\bibfnamefont {P.~H.}\ \bibnamefont {Heenen}},\ and\ \bibinfo
  {author} {\bibfnamefont {J.}~\bibnamefont {Meyer}},\ }\bibfield  {title}
  {\bibinfo {title} {{Analysis of the generator coordinate method in a study of
  shape isomerism in 194 Hg}},\ }\href
  {https://doi.org/10.1016/0375-9474(90)90062-Q} {\bibfield  {journal}
  {\bibinfo  {journal} {Nucl. Phys. A}\ }\textbf {\bibinfo {volume} {510}},\
  \bibinfo {pages} {466} (\bibinfo {year} {1990})}\BibitemShut {NoStop}%
\bibitem [{\citenamefont {Hara}\ \emph {et~al.}(1982)\citenamefont {Hara},
  \citenamefont {Hayashi},\ and\ \citenamefont {Ring}}]{Hara:1982egw}%
  \BibitemOpen
  \bibfield  {author} {\bibinfo {author} {\bibfnamefont {K.}~\bibnamefont
  {Hara}}, \bibinfo {author} {\bibfnamefont {A.}~\bibnamefont {Hayashi}},\ and\
  \bibinfo {author} {\bibfnamefont {P.}~\bibnamefont {Ring}},\ }\bibfield
  {title} {\bibinfo {title} {{Exact angular momentum projection of cranked
  Hartree-Fock-Bogoliubov wave functions}},\ }\href
  {https://doi.org/10.1016/0375-9474(82)90486-9} {\bibfield  {journal}
  {\bibinfo  {journal} {Nucl. Phys. A}\ }\textbf {\bibinfo {volume} {385}},\
  \bibinfo {pages} {14} (\bibinfo {year} {1982})}\BibitemShut {NoStop}%
\end{thebibliography}

\end{document}